\documentclass[12pt,oneside,reqno,intlimits]{amsart}
\usepackage{setspace}
\usepackage{graphicx}
\usepackage{amssymb}

\usepackage{subcaption}

\usepackage{pdflscape}
\usepackage{fullpage}
\usepackage{commath}
\usepackage{float}
\usepackage{multirow}
\usepackage{verbatim}
\usepackage{changepage}

\usepackage{algpseudocode}
\usepackage{algorithm}
\usepackage{url}
\urldef\myurl\url{ }

\usepackage{wrapfig, blindtext}
\usepackage{hhline}
\usepackage{enumitem}
\usepackage[table]{xcolor}
\definecolor{lightgray}{gray}{0.9}

\usepackage{newcent}
\usepackage{cases}
\usepackage{hyperref}
\usepackage{setspace}
\usepackage{wasysym}
\usepackage{braket}
\usepackage[utf8]{inputenc}
\usepackage[T1]{fontenc}
\usepackage{todo}
\newcommand{\cl}[0]{\mathrm{c}}

\newcommand*{\horzbar}{\rule[.5ex]{2.5ex}{0.5pt}}
\usepackage[foot]{amsaddr}

\title[TDOA Source Localization]{Time Difference of Arrival Source Localization:\\Exact Linear Solutions for the General 3D Problem}
\author{Niraj K. Inamdar}
\email{nki@alum.mit.edu}
\begin{document} 
\maketitle

\section{Introduction}
The time difference of arrival (TDOA) problem admits exact, purely algebraic solutions for the situation in which there are 4 and 5 sensors and a single source whose position is to be determined in 3 dimensions. The solutions are exact in the sense that there is no least squares operation (i.e., projection) involved in the solution.  The solutions involve no linearization or iteration, and are algebraically transparent via vector algebra in Cartesian coordinates. The solution with 5 sensors requires no resolution of sign ambiguities; the solution with 4 sensors requires resolution of one sign ambiguity. Solutions are effected using only TDOA and not, e.g., frequency difference of arrival (FDOA) or angle of arrival (AOA) \cite{Ho}.

We note previous work towards achieving closed-form solutions to the TDOA problem in 2 dimensions (\cite{hubavcek,le}) and 3 dimensions (\cite{spencer}). Our work differs from the former in that it is fully general in 3 dimensions. It differs from the latter in that we demonstrate a purely linear algebraic solution in Cartesian coordinates without auxiliary variables and show clearly how the requirement for ambiguity resolution arises for the 4-sensor case but does not appear in the 5-sensor case. 

We first present the 5-sensor solution (Section \ref{sec:FiveSensor}) and then follow with the 4-sensor scenario (Section \ref{sec:FourSensor}). Numerical experiments (Section \ref{sec:NumExp}) are presented showing the performance of the calculations in the case of no noise, before closing with conclusions (Section \ref{sec:Conc}). Performance of the calculations below is exact within numerical error, and in the small fraction of cases in which source localization does not occur, it is driven by misidentification in resolution of sign ambiguity without priors. 

Our results are suitable for self-localization applications such as simultaneous localization and mapping (SLAM) and precision navigation and timing (PNT, through, e.g., GPS), as well electromagnetic and acoustic signal source geolocation problems. For their speed, exactness, and linearity, we believe the calculations below have substantial practical utility.

\section{The TDOA Solution for 5 Sensors in 3 Dimensions}\label{sec:FiveSensor}
First we consider the 5-sensor scenario. Let the sensors have positions given by $\textbf{r}'_k$, $k = 1,...,5$ and the source position, to be solved for, be given by $\textbf{r}'_S$. If $\textbf{r}'_1$ is set to be the origin of the coordinate system \cite{Smith}, then $\textbf{r}_k \equiv \textbf{r}'_k - \textbf{r}'_1$, $k = 2,...,5$, and $\textbf{r}_1 = 0$.

The ranges from the source to the sensors, $\rho_k = |\textbf{r}_k - \textbf{r}_S|$, are unknowns, while the quantities $\rho_k - \rho_j \equiv \delta_{kj}$ are measured/inferred quantities based on time arrival differences of some waveform. That is, for measurement times $t_k$ and $t_j$ at the $k\textsuperscript{th}$ and $j\textsuperscript{th}$ sensor respectively, $\rho_k - \rho_j = \cl(t_k - t_j)$ where $\cl$ is the speed of light and the $t$s are corrected times of arrival at each sensor. Since $\textbf{r}_1 = 0$, $\rho_1 = \sqrt{\textbf{r}_S^T \textbf{r}_S} = |\textbf{r}_S|$.

We have for $\delta_{k1}$, $k>2$,
\begin{eqnarray}
\delta_{k1} = \rho_k - \rho_1 &= \sqrt{\textbf{r}_k^T \textbf{r}_k - 2 \textbf{r}_k^T\textbf{r}_S + \textbf{r}_S^T \textbf{r}_S} - \sqrt{\textbf{r}_S^T \textbf{r}_S}.
\end{eqnarray}
Upon rearranging and squaring, we get an expression for $\rho_1$ that is linear in the source position $\textbf{r}_S$:
\begin{eqnarray}\label{eq:rhoDef}
\rho_1 = \frac{\textbf{r}_k^T \textbf{r}_k - 2\textbf{r}_k^T \textbf{r}_S - \delta_{k1}^2}{2\delta_{k1}}.
\end{eqnarray}
The expression for $\rho_1$ can be substituted into any expression for $\delta_{j1}$ ($j\neq 1, j \neq k$) to yield an expression for $\textbf{r}_S$:
\begin{eqnarray}\label{eq:rhosubSol}
2\left(\textbf{r}_k^{T} - \frac{\delta_{k1}}{\delta_{j1}}\textbf{r}_j^{T}\right)\textbf{r}_S = -\left(\delta_{k1}^2 - \frac{\delta_{k1}}{\delta_{j1}}\delta_{j1}^2\right) + \left[ \textbf{r}_k^{T}\textbf{r}_k - \frac{\delta_{k1}}{\delta_{j1}} \left( \textbf{r}_j^{T}\textbf{r}_j \right)\right].
\end{eqnarray}
The vector transpose multiplying $\textbf{r}_S$ on the left hand side of Eq. \eqref{eq:rhosubSol} is built up into a matrix $\textbf{B}$, while the scalars on the right hand side are built up into a vector $\textbf{x}$ so that $\textbf{B}\textbf{r}_S = \textbf{x}$. Since we are solving for a $3\times 1$ array, we need 3 pairs of indices $(k,j)$ that will maintain the full rank of $\textbf{B}$. We choose $(k,j)  = \{(3,2),(4,3),(5,4)\} $, so that explicitly, we have
\begin{eqnarray}
\textbf{B} = \left[ 
\begin{array}{c}
\horzbar~2\left(\textbf{r}_3^{T} - \displaystyle\frac{\delta_{31}}{\delta_{21}}\textbf{r}_2^{T}\right)\horzbar \\
\horzbar~2\left(\textbf{r}_4^{T} - \displaystyle\frac{\delta_{41}}{\delta_{31}}\textbf{r}_3^{T}\right)\horzbar \\
\horzbar~2\left(\textbf{r}_5^{T} - \displaystyle\frac{\delta_{51}}{\delta_{41}}\textbf{r}_4^{T}\right)\horzbar 
\end{array}
\right]
\end{eqnarray}
and 
\begin{eqnarray}
\textbf{x} = \left[ 
\begin{array}{c}
-\left(\delta_{31}^2 - \displaystyle\frac{\delta_{31}}{\delta_{21}}\delta_{21}^2\right) + \left[ \textbf{r}_3^{T}\textbf{r}_3 - \displaystyle\frac{\delta_{31}}{\delta_{21}} \left( \textbf{r}_2^{T}\textbf{r}_2 \right)\right]\\
-\left(\delta_{41}^2 - \displaystyle\frac{\delta_{41}}{\delta_{31}}\delta_{31}^2\right) + \left[ \textbf{r}_4^{T}\textbf{r}_4 - \displaystyle\frac{\delta_{41}}{\delta_{31}} \left( \textbf{r}_3^{T}\textbf{r}_3 \right)\right]\\
-\left(\delta_{51}^2 - \displaystyle\frac{\delta_{51}}{\delta_{41}}\delta_{41}^2\right) + \left[ \textbf{r}_5^{T}\textbf{r}_5 - \displaystyle\frac{\delta_{51}}{\delta_{41}} \left( \textbf{r}_4^{T}\textbf{r}_4 \right)\right]
\end{array}
\right].
\end{eqnarray}
Thus, 
\begin{eqnarray}
\textbf{r}_S = \textbf{B}^{-1}\textbf{x}.
\end{eqnarray}
The solution is exact (so long as sensor geometry ensures \textbf{B} has rank 3) and requires no resolution of sign ambiguities. Note that since the coordinates have been referenced to $\textbf{r}'_1$, we have the actual source location $\textbf{r}'_S$ as 
\begin{eqnarray}\label{eq:actSol}
\textbf{r}'_S = \textbf{r}_S + \textbf{r}'_1.
\end{eqnarray}

\section{The TDOA Solution for 4 Sensors in 3 Dimensions}\label{sec:FourSensor}
The 4-sensor solution has its starting point in Eq. \eqref{eq:rhoDef}. Taking $k = 2,3,4$, we have 
\begin{eqnarray}
2\delta_{21} \rho_1 &=& \textbf{r}_2^T \textbf{r}_2 - 2\textbf{r}_2^T \textbf{r}_S - \delta_{21}^2\\
2\delta_{31} \rho_1 &=& \textbf{r}_3^T \textbf{r}_3 - 2\textbf{r}_3^T \textbf{r}_S - \delta_{31}^2\\
2\delta_{41} \rho_1 &=& \textbf{r}_4^T \textbf{r}_4 - 2\textbf{r}_4^T \textbf{r}_S - \delta_{41}^2,
\end{eqnarray}
or, defining
\begin{eqnarray}
\textbf{z} = 
\left[
\begin{array}{c}
2\delta_{21} \\
2\delta_{31} \\
2\delta_{41} 
\end{array}
\right],~~\textbf{y} = 
\left[
\begin{array}{c}
\textbf{r}_2^T \textbf{r}_2 - \delta_{21}^2 \\
\textbf{r}_3^T \textbf{r}_3 - \delta_{31}^2 \\
\textbf{r}_4^T \textbf{r}_4 - \delta_{41}^2
\end{array}
\right],~~\textbf{C} = 
\left[
\begin{array}{c}
\horzbar  - 2\textbf{r}_2^T \horzbar\\
\horzbar  - 2\textbf{r}_3^T \horzbar\\
\horzbar  - 2\textbf{r}_4^T \horzbar
\end{array}
\right],
\end{eqnarray}
we have equivalently 
\begin{eqnarray}
\rho_1 \textbf{z} = \textbf{y} + \textbf{C}\textbf{r}_S.
\end{eqnarray}
If the sensor geometry ensures $\textbf{C}$ has rank 3, then we have
\begin{eqnarray}\label{eq:r_SSol}
\textbf{r}_S = \rho_1 \textbf{C}^{-1}\textbf{z} - \textbf{C}^{-1}\textbf{y}.
\end{eqnarray}
Define for convenience $\boldsymbol{\xi} \equiv \textbf{C}^{-1}\textbf{z}$ and $\boldsymbol{\eta} \equiv \textbf{C}^{-1}\textbf{y}$. Then using $\textbf{r}_S^T\textbf{r}_S = \rho_1^2$ gives a quadratic equation for $\rho_1$
\begin{eqnarray}
\rho_1^2 \left(\boldsymbol{\xi}^T\boldsymbol{\xi} - 1\right) - 2\boldsymbol{\xi}^T\boldsymbol{\eta}\rho_1 + \boldsymbol{\eta}^T\boldsymbol{\eta}=0,
\end{eqnarray}
with a solution 
\begin{eqnarray}\label{eq:rhoSolution}
\rho_1 = \frac{\boldsymbol{\xi}^T\boldsymbol{\eta} \pm \sqrt{(\boldsymbol{\xi}^T\boldsymbol{\eta})^2 - \left(\boldsymbol{\xi}^T\boldsymbol{\xi} - 1\right)\boldsymbol{\eta}^T\boldsymbol{\eta}}}{\left(\boldsymbol{\xi}^T\boldsymbol{\xi} - 1\right)}.
\end{eqnarray}
Once $\rho_1$ is solved for, the solution for $\textbf{r}_S$ is effected through Eq. \eqref{eq:r_SSol}, and the actual source position via Eq.\eqref{eq:actSol}. 

The sign ambiguity in Eq. \eqref{eq:rhoSolution} can usually be resolved by taking the two source positions corresponding to the two $\rho_1$ solutions ($\textbf{r}_S(\rho_1^{(1)})$ and $\textbf{r}_S(\rho_1^{(2)})$, say), recalculating the corresponding $\rho_k - \rho_1$, and determining the one of two solutions that minimizes $\sum_{k = 2}^{4}\left[(\rho_k - \rho_1) - \delta_{k1}\right]^2$. The presence of priors (positions constrained, e.g., by fixed beam widths) can also be used to constrain the solution. 

\section{Numerical Experiments}\label{sec:NumExp}
The calculations above have been numerically tested with generally excellent, exact performance. Suppose the source location is varied in fully 3-dimensional space ($\pm x$, $\pm y$, and $\pm z$) over 1,000 Monte Carlo instances, and the locations of up to 5 sensors are also varied in 3-dimensional space. The sensor positions $\textbf{r}_k'$ ($k = 1,...,5$) and source position $\textbf{r}_S' $ are drawn from uniform distributions $\mathcal{U}(0,1)$. The source positions are multiplied by a scaling factor $\mathtt{SOURCE\_SCALE}$ relative to the sensor locations:
\begin{eqnarray}
\textbf{r}_k' &\sim& \left\{ \left[\mathcal{U}(0,1),\mathcal{U}(0,1),\mathcal{U}(0,1)\right]^T - 0.5 \right\}, k = 1,...,5\\
\textbf{r}_S' &\sim& \mathtt{SOURCE\_SCALE}\times\left\{ \left[\mathcal{U}(0,1),\mathcal{U}(0,1),\mathcal{U}(0,1)\right]^T - 0.5 \right\}.
\end{eqnarray}

In each experiment, the inferred source position $\tilde{\textbf{r}}_S'$ is compared to the truth source position $\textbf{r}_S'$ with no measurement noise. For each Monte Carlo instance, if the relative error $|\tilde{\textbf{r}}_S' - \textbf{r}_S|/|\textbf{r}_S|$ is less than some threshold $\mathcal{T}$, the calculation is considered a success. For a given $\mathtt{SOURCE\_SCALE}$, the fraction of successes over all 1,000 Monte Carlo runs is calculated. In Fig. \ref{fig:NumExps} below, we show results from numerical experiments for cases with $\mathcal{T} = 10^{-6},10^{-3}$ and $\mathtt{SOURCE\_SCALE} \in [10^{-6},1]$ . 

The very small fraction of errors that does exist is driven by numerical matrix inversion errors for cases in which sensor positions have poor geometric diversity and when the source positions are highly constrained compared to the sensors. For the 4-sensor solution, errors that exist for $\mathtt{SOURCE\_SCALE}$ have been verified to be driven entirely by the incorrect choice in $\rho_1$ when solving Eq. \eqref{eq:rhoSolution}. Hence, in all cases (4- or 5-sensor), the inferred solution for no noise is within numerical error of the exact value, and even in incorrect inferences in the limiting 4-sensor case, the correct solution is still one of two tested values that may potentially be easily selected based on priors. Depending on the application, the existence of this error may therefore have negligible practical impact.

\section{Conclusions}\label{sec:Conc}
We have presented here fully algebraic solutions to the TDOA problem. For the case of 5 sensors, the solution method requires only inverting a set of 3 linear equations. For the case of 4 sensors, the solution method requires inverting a set of 3 linear equations and solving one quadratic equation. Numerical experiments have been carried out using the techniques derived for varying sensor and source positions, each over 1,000 Monte Carlo instances with excellent, exact performance in the noiseless case. To the best of our knowledge, this is the first presentation of exact, linear solutions to the TDOA problem for 4- or 5-sensor scenarios in  3 dimensions. 

\begin{figure}[h!]
\begin{center}
 \includegraphics[width=\textwidth]{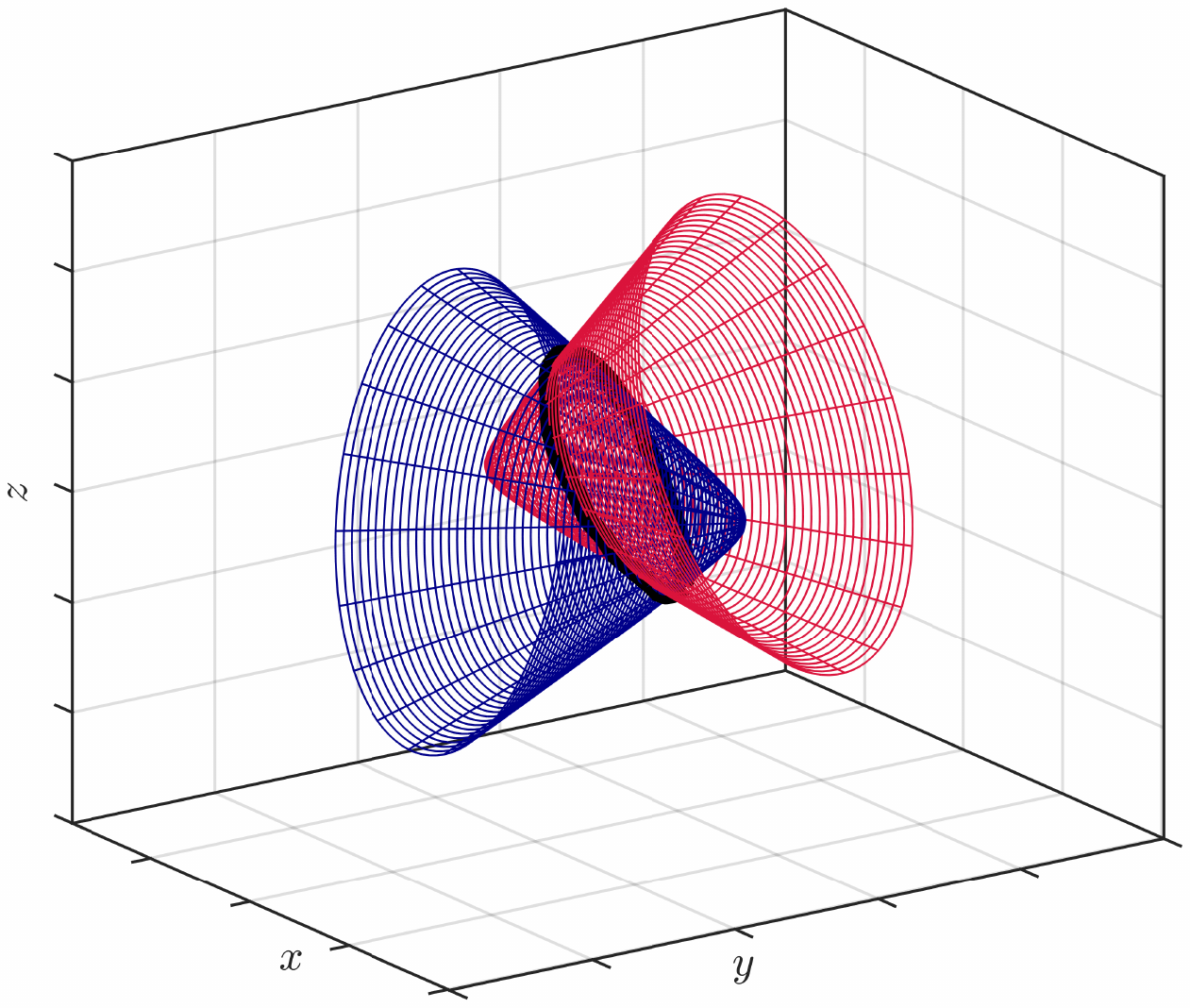}
          \caption{Intersection of two hyperboloids indicated in black.}
          \label{fig:intHyper}
\end{center}
\end{figure}

That 4 sensors require ambiguity resolution while 5 sensors do not is perhaps evident from geometrical considerations, although the presence of a purely linear algebraic solution to the 5-sensor case might be somewhat unexpected. For 2 sensors in 3 dimensions, the TDOA problem possess a degeneracy about the surface of a hyperboloid. Introducing a third sensor reduces this degeneracy to a line (the intersection between two hyperboloids; Fig. \ref{fig:intHyper}), while a fourth sensor reduces the source location to up to two points, depending on the shape of the line and the new hyperboloid. The introduction of a fifth sensor ensures that if there are two possible solutions from the 4-sensor case, only the viable one is selected. 

The fact that our solutions are linear make them amenable to rapid, on-platform calculation. The lack of need for iteration obviates concerns about convergence. We believe there is substantial practical utility in implementing these solutions for a range of localization and tracking problems.

\begin{figure}[h!]
    \centering
      \begin{subfigure}{0.67\textwidth}
        \includegraphics[width=\textwidth]{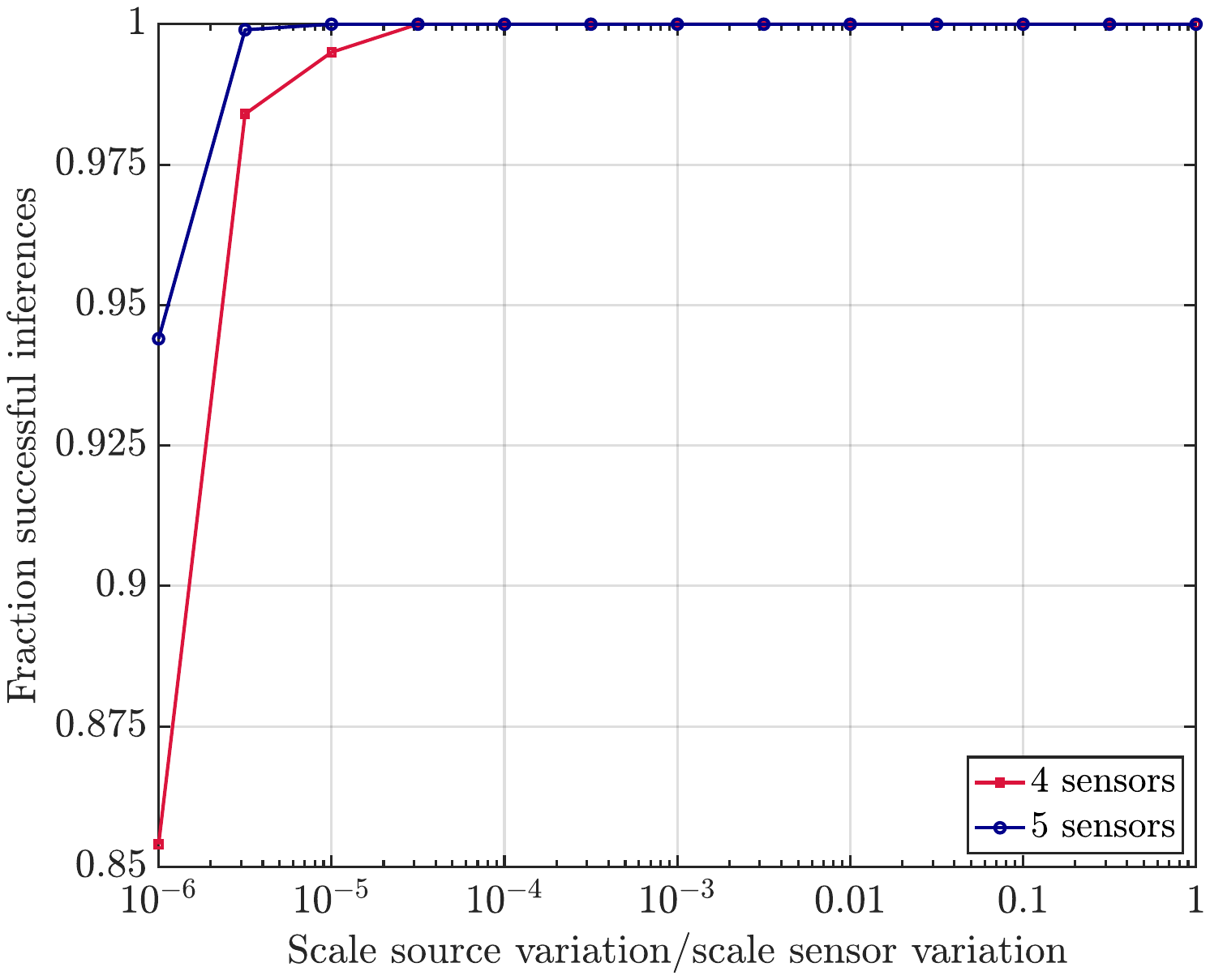}
          \caption{Source localization success fraction as a function of position variation. Relative position accuracy threshold of $\mathcal{T} = 10^{-6}$ for identification success.}
          \label{fig:NumExps1}
      \end{subfigure}
      \\
      \begin{subfigure}{0.67\textwidth}
        \includegraphics[width=\textwidth]{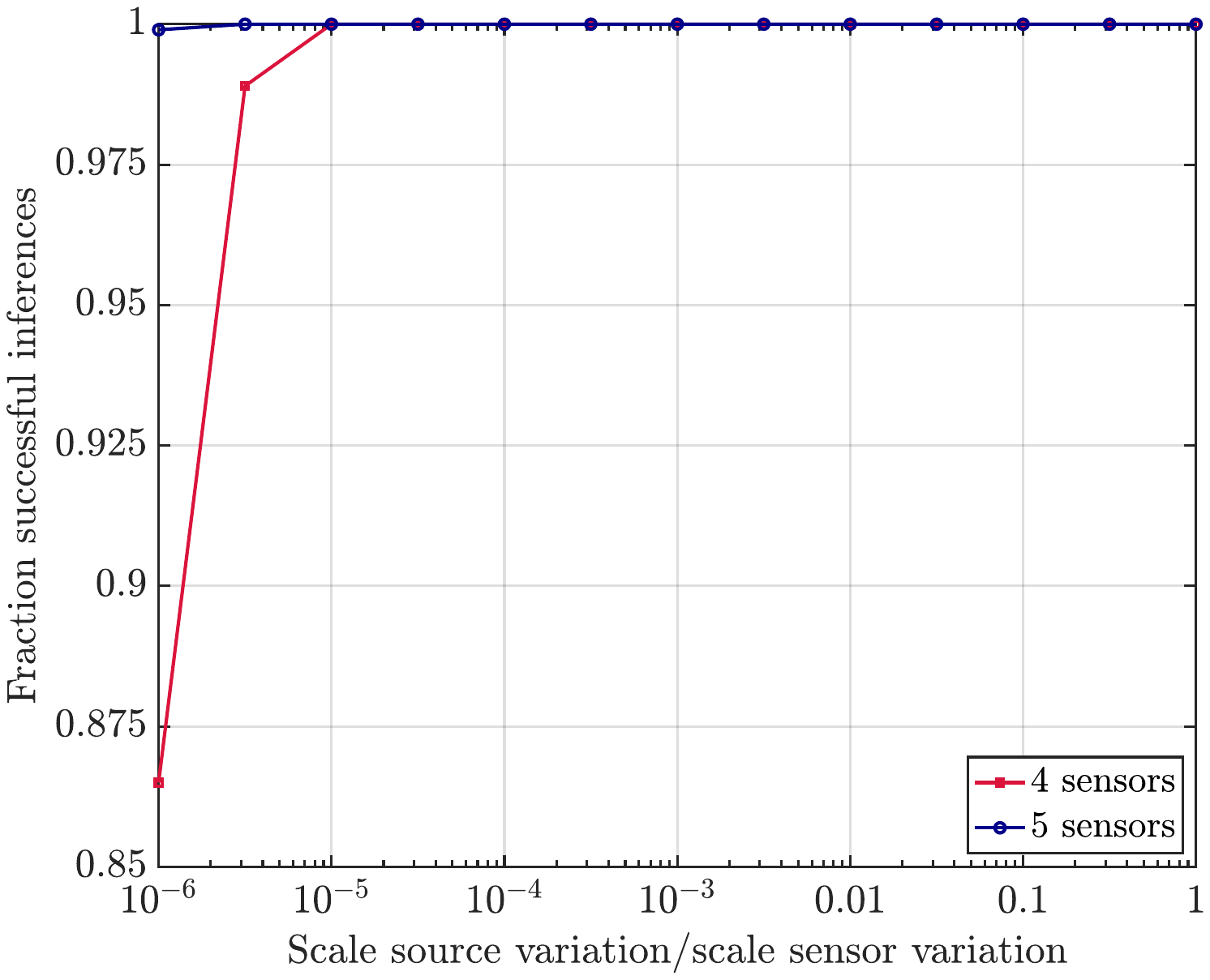}
          \caption{Source localization success fraction as a function of position variation. Relative position accuracy threshold of $\mathcal{T} = 10^{-3}$ for identification success.}
          \label{fig:NumExps2}
      \end{subfigure}
\caption{
\label{fig:NumExps}%
Results from numerical experiments showing performance of TDOA source localization over 1,000 Monte Carlo instances. Source positions are calculated using the methods outlined in the text.}
\end{figure}

\newpage
\bibliographystyle{unsrt}
\bibliography{TDOA_refs}
 
\end{document}